# Spiking ruby revisited: self-induced periodic spiking oscillations leading to chaotic state in a Cr:Al$_2$O$_3$ laser with cw 532-nm pumping

KENJU OTSUKA,[1*] AND SEIICHI SUDO[2]

[1]*TS$^3$L Research, 126-7 Yamaguchi, Tokorozawa, Saitama 359-1145 Japan*
[2]*Department of Physics, Tokyo City University, Tokyo 158-8557 Japan*
*[\*kenju.otsuka@gmailcom](mailto:kenju.otsuka@gmailcom)*

**Abstract:** This paper reexamines a 60-year-old mystery of spiking behavior in ruby lasers with a cw 532-nm pump, paying special attention to mode matching between the pump and lasing beam within a ruby crystal placed in a semi-confocal laser cavity. Periodic spiking oscillations were observed in a limited pump power regime, where spikes obeying the generic asymmetric hyperbolic function appeared at a repetition rate around 50 kHz and with a 130-150 ns width and 0.1-0.6 µJ energy depending on the pump power. The physics of the spiking behavior based on Kleinman's mechanical approach and a plausible interpretation for the periodic spiking oscillation in terms of self-induced mode matching between the pump and laser beams through the self-induced Kerr-lens effect are addressed. The statistical nature inherent to spiking and the associated self-organized critical behavior in the quasi-periodic spiking oscillations as well as chaotic states occurring outside the periodic spiking regime are clarified from a nonlinear dynamics point of view and intriguing statistical properties inherent to chaotic oscillations in usual solid-state lasers subjected to external modulations are shown to be present in the self-induced instabilities of the ruby laser

## 1. Introduction

The tremendous progress that has been witnessed in the laser sciences and optical technologies all started with T. Maiman's ruby laser [1]. A few years after the development of the ruby laser, a certain problem was nearly forgotten until the advent of the Ar laser pumped cw ruby laser [2-6]. That is, when this laser was first demonstrated, its output was regarded as being completely stable under cw pumping at helium or nitrogen temperatures [2, 4]. Later, other investigators disagreed and reported dynamical instabilities [3, 5, 6]. It has become clear since then that ruby lasers can operate under both conditions. Not only can the laser be completely stable, but it can also operate in two instability conditions of chaotic and regular pulsing output [7]. The repetitive self-switching as well as related undamped spiking phenomena still remain as interesting unsolved mysteries in ruby and other lasers. There is to date still no definite understanding of the physical mechanism driving the instability.

On the other hand, two-orders of spectral narrowing of the fluorescence line width is required by cooling the rod to liquid nitrogen temperatures for cw argon ion laser pumping at 514.5 nm. Recently, two research groups have realized room-temperature cw ruby lasers by pumping at the optimum of the absorption profile at 405 nm of GaN laser diodes [8, 9], where both employed ruby crystals with a Cr$_2$O$_3$ concentration of 0.05 % and a c-axis at 90º relative to the rod axis. These cw ruby laser oscillations with GaN laser-diode pumping raise intriguing questions as to whether room-temperature LD pumping at 405 nm can solve the over 60-year-old mystery of the spiking behaviors in ruby lasers.

Here, we performed systematic experiments towards room-temperature cw operation of ruby lasers pumped with a frequency-doubled Nd:YAG laser at 532 nm. Unfortunately, non-spiking cw oscillations were not achieved, while self-induced periodic spiking oscillations leading to quasi-periodic spiking and chaotic states were observed. The physics of the spiking behavior based on Kleinman's mechanical approach and a plausible interpretation for the periodic spiking oscillation in terms of self-induced mode matching between pump and laser beams are addressed.

## 2. Experimental results

### 2.1 Experimental apparatus and input-output characteristics

The experimental setup is shown in Fig. 1(a). A $TEM_{00}$ second harmonic output (wavelength: 532 nm) of an LD-pumped linearly-polarized $TEM_{00}$ mode Nd:YAG laser (CNI MG-F-532; $TEM_{00}$ mode, $M^2$ = 1.2), whose polarization direction was set perpendicular to the ruby c-axis by a half-wave plate, was focused on a 7-mm-thick, 10-mm-diameter a-cut ruby single crystal ($Cr^{3+}$:$Al_2O_3$) with a 0.03 wt% Cr concentration (SHINKOSHA Co., Ltd.), while the absorption coefficient at 532 nm was measured to be a = 2.1 $cm^{-1}$. The end surface was directly coated with a dielectric mirror $M_1$ (transmission at 532 nm: 88%, reflectance at 694 nm: 99.8%). A concave mirror $M_2$ (reflectance at 694 nm: 99%; radius of curvature, $R_2$ = 5 cm) was placed nearly 2.5 cm in optical length apart from the mirror $M_1$ to form the most stable semi-confocal laser cavity. The pump beam spot size averaged over the crystal was made close to the semi-confocal cavity spot size (around 70 μm) by controlling the pump beam diameter at the lens of focal length f = 15 cm for decreasing the threshold pump power. The lasing pattern was measured by a mode profiler and the output was detected by a Si photo-diode (New Focus 1801M, 125 MHz bandwidth) followed by a digital phosphor oscilloscope (Tektronix DPO 2024, 200-MHz bandwidth). The pumping intensity fluctuation measured on the same time scale as the following experiment was at most ±1% in the entire pump power regime.

The input-output characteristics are shown in Fig. 1(b), where the far-field lasing pattern exhibited a $TEM_{00}$ transverse mode as shown in the inset. The dynamic behavior was found to change with increasing pump power, as depicted in Fig. 1(b).

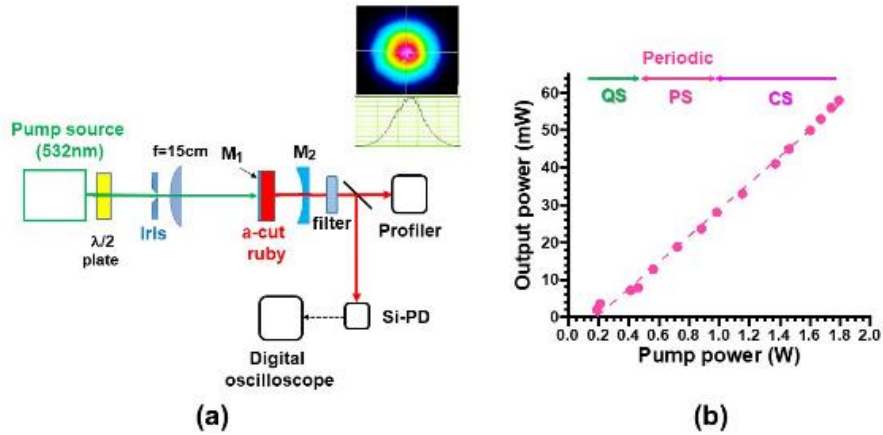

Fig. 1. (a) Experimental apparatus. (b) Input-output characteristics. QS: quasi-periodic spiking regime, PS: periodic spiking regime, CS: chaotic state.

### 2.2 Dynamic behavior

The present ruby laser exhibited multi-longitudinal mode operation separated by $\Delta f$ = c/2nl = 12.1 GHz (i.e., $\Delta\lambda = \lambda^2/2nl$ = 0.2Å in spectrometer measurement), which resulted from the wavelength selection effect of the 7-mm thick active "crystal etalon" (c: velocity of light, refractive index n = 1.77, crystal length l = 7mm) [10]. Such multimode oscillations in homogeneously broadened solid-state lasers, e. g. ruby lasers, have been explained in terms of the spatial hole-burning effect of population inversions by C. L. Tang, *et. al* in 1963 [11]. In addition, individual modes have been proven to exhibit self-organized collective behaviors such that the total output behaves just like a single mode laser through the cross-saturation of population inversions in the vicinity of stationary state and in large signal regimes including chaos [12-14].

In the present experiment, dynamical properties were measured for the total output power focused on a Si photo-diode without using a spectrometer. Therefore, the following dynamical behaviors are considered to represent the generic properties of a single-mode laser. Periodic spiking oscillations appeared in the limited pump power regime, PS, while the periodicity was undermined in the lower pump region, QS, featuring quasi-periodicities and a chaotic state appeared in the higher pump regime, CS, above the PS regime as depicted in Fig. 1(b). An example of a periodic spiking oscillation is shown in Fig. 2(a), together with the corresponding power spectrum and a map of the phase-space attractor [7]. As for the mapping, we used a total data length of $T_{total} = \Delta\tau \times 10^5$, which was restricted by the digital oscilloscope, by setting the measurement time interval to $\Delta\tau$ = 4 ns to reproduce narrow spiking waveforms accurately.

The pump-dependent repetition frequency, spike pulse width, and associated spike-energy are shown in Fig. 2 (b), where the peak output power was calibrated from the averaged output power shown in Fig. 1(b). The peak power of spikes within the cavity reached 100~400 W, reflecting extremely low duty cycles of 0.6~0.75%.

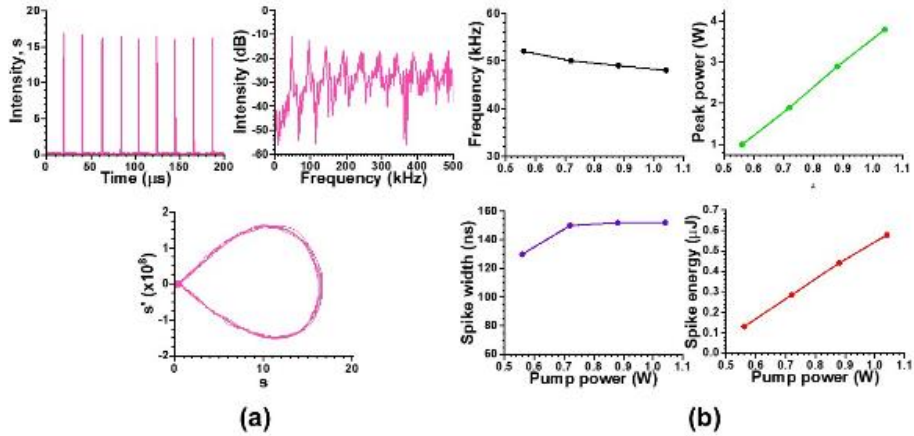

Fig. 2. Periodic spiking oscillation. (a) Waveform, power spectrum, and Poincaré section at pump power of P = 1040 mW. (b) Pump-dependent repetition frequency, peak power, pulse width and energy of spiking pulse.

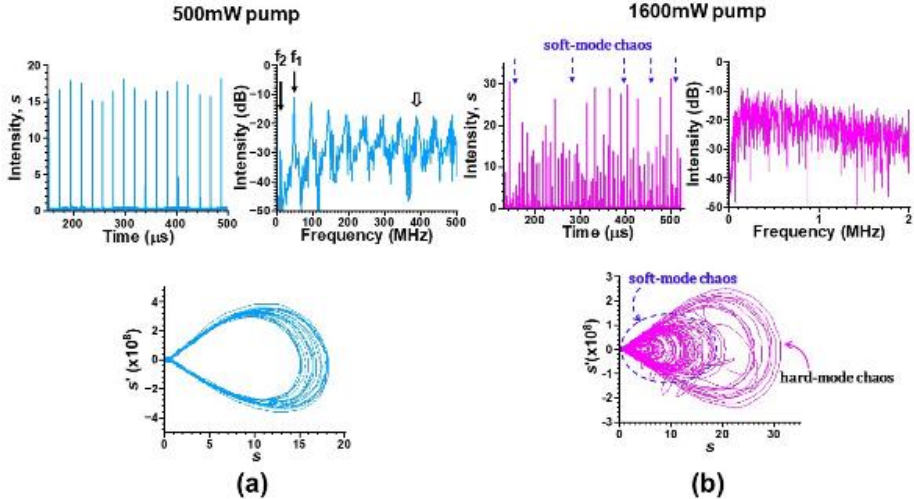

Fig. 3. (a) Quasi-periodic spiking oscillation featuring two competing frequencies, $f_1$ = 9.5 kHz and $f_2$ = 48.75 kHz, whose higher harmonic components merge in the vicinity of 390 kHz. (b) Chaotic state featuring alternating soft and hard-mode chaos over time which will discussed again in section 5.

Results for the quasi-periodic spiking and chaotic state are summarized in Fig. 3. They capture the qualitative difference in the phase space trajectories such that isomorphic "multiple loops" appear in the quasiperiodic spiking, while "chaotic loops" featuring small and large orbits, which indicate alternate appearances of soft and hard-mode chaos (see section **5**), appear in the chaotic state.

## 3. Statistical nature of spiking pulse

To provide physical insight into the periodic excitation of the spiking oscillations, let us introduce the Toda potential for the laser rate equations. According to Kleiman's mechanical approach relevant to four-level as well as three-level systems like a ruby laser, which are governed by standard rate equations for population inversions and photon density [15,16], the dynamics of the photon density can be understood in analogy with the motion of a particle in the following laser Toda potential, $V$, with a logarithmic transformation for the photon density as $u(t) \equiv \ln s(t)$ [17]:

$$d^2 u/dt^2 + \kappa (du/dt) + \partial V/\partial u = F_D(t), \tag{1}$$

$$V = K[(e^u - 1) - (w - 1)u], \tag{2}$$

$$\kappa = 1 + e^u, \tag{3}$$

where $w = P/P_{th}$ (P: pump power, $P_{th}$: threshold pump power), $K = \tau/\tau_p$ ($\tau$: fluorescence lifetime, $\tau_p$: photon lifetime), $F_D$ is a driving force to the particle, while the effect of spontaneous emission is neglected for the sake of brevity. Here, the damping rate, $\kappa$, increases as the photon density, $s(t) = e^{u(t)}$, increases, whereas it does not depend on u(t) in the original Toda oscillator [18]. The pump-dependent laser Toda potentials, $V(u)$, are shown in Fig. 4(a).

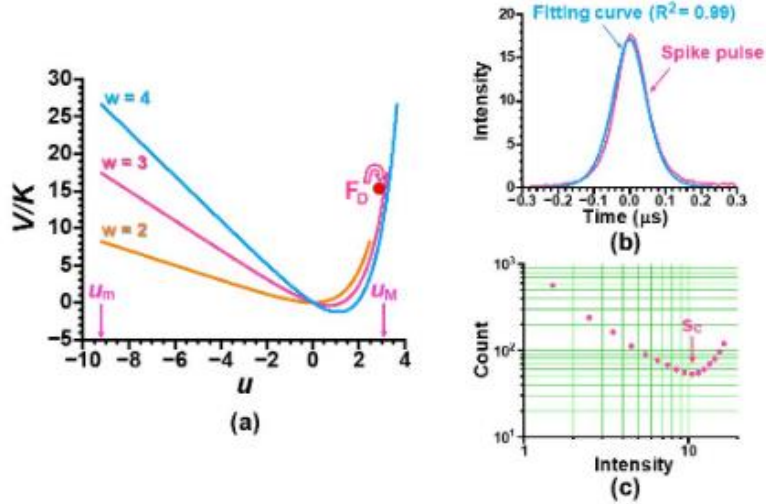

Fig. 4. (a) Pump-dependent laser Toda potentials. $w = P/P_{th}$. (b) Zoomed-in view of spike waveform, together with the hyperbolic fitting curve. (c) Intensity probability distribution for periodic spiking oscillations shown in Fig. 2(a).

The particle moves within a highly asymmetric laser Toda potential with a damping rate, $\kappa$. Without a driving force, $F_D$, in Fig. 4(a), the particle approaches the ground state, exhibiting damped relaxation oscillations. Hamiltonian motion around the ground state, i.e., periodic relaxation oscillations (soft mode), is established if a periodic driving force is applied at the relaxation oscillation frequency, $f_{RO} = (1/2\pi)[(w - 1)/\tau\tau_p]^{1/2}$, such that the damping force, $\kappa(du/dt)$, balances the periodic driving force, $F_D$ [17]. In addition to the soft mode, a spike-like waveform builds up within the asymmetric potential by tuning the strength and frequency of the driving force in the large signal regime. In fact, periodic spiking oscillations (hard-mode) were realized in semiconductor lasers through the use of deep injection current modulation [19] and in solid-state lasers through the use of a deep pump [20] or loss modulation [17] at $f_{SP}$ (< $f_{RO}$).

Let us examine whether the self-induced periodic spiking oscillations shown in Fig. 2(a) obey the generic dynamic properties of spiking oscillations reported so far. The spike-pulse waveform is confirmed to be well fitted by the following hyperbolic function, as in ref. [17]:

$$s(t) = s_p \text{sech}^2 \left( \sqrt{\frac{s_p}{2\tau\tau_p}} (t - t_0) \right), \tag{4}$$

Here, $t_o$ is the time at which the peak photon number occurs. A magnified view of a single spike-pulse waveform in the ruby laser and the hyperbolic fitting curve are shown in Fig. 4(b), assuming $\tau = 3.4$ ms and $\tau_p = 2$ ns. A large coefficient of determination of $R^2 = 0.99$ is attained in this case. The intensity probability distribution $P(s)$ for the periodic spiking oscillation corresponding to Fig. 2(a) reaches a minimum at the inflection point, $s_c = s_p \text{sech}^2 \left( \text{arctanh} \frac{1}{\sqrt{3}} \right) = \frac{2}{3} s_p$, as shown in Fig. 4(c), where $P(s)$ increases monotonically in the region $s > s_c$ [17]

## 4. Transverse mode formation within ruby crystal

### 4.1 Thermal and Kerr lens effects

A peculiar statistical property relevant to the laser's spiking oscillations shown in the previous section strongly suggests that the appropriate self-induced driving force forms so as to cancel the damping force to thereby establish periodic motions preserving the potential energies, $V(u_M) = V(u_m) = E$, over time, as in Fig. 4(a). Ever since the first report in 1968, self-induced pulsations in ruby lasers have drawn interest and have been given a number of plausible explanations in terms of "self-Q-switching" (SQS), which is based on the saturable absorber effect inherent to impurities or ground-state reabsorption effect in the unpumped region, as well as cavity misalignment [6, 21-23]. Here, we investigate the mechanism of self-induced periodic spiking oscillations in the present ruby laser by focusing on the matching between the pump and lasing beam profiles within the ruby crystal.

First of all, let us examine the thermal lens effect which modifies the transverse eigenmode to be formed in the laser resonator. The focal length of the thermal lens is given by

$$f_T = \frac{\pi w_p^2 K_T}{2Q\left(\frac{dn}{dT}\right)} \tag{5}$$

$$Q = P_{abs}\left(1 - \frac{v_l}{v_p}\right), \tag{6}$$

where $P_{abs}$ is absorbed pump power, $v_p$ and $v_l$ are pump and laser optical frequencies, $w_p$ is the average pump spot size in the crystal, and $K_T$ is thermal conductivity [7]. The estimated focal length within the periodic spiking regime is shown by the red curve in Fig. 5(b) as a function of the pump power in the periodic spiking regime in Fig. 2(b), assuming $K_T = 0.092$ cal/cm-°C-sec and $dn/dT = 1.2 \times 10^{-5}$/°C for ruby crystal.

In the case of a ruby laser, on the other hand, the complex nonlinear refractive index originates from the light-induced population changes in the excited and ground states of the $Cr^{3+}$ ion. Indeed, a huge *resonance-enhanced* nonlinear refractive index of $n_2 = 1.25 \times 10^{-12}$ m²/W was reported in an experiment based on the differential interferometric technique using an argon ion laser for the absorption band at 514.5 nm [24], which resulted in self-induced superluminal and subluminal group velocity propagation in the pink ruby [25, 26]. The optical bistability was demonstrated in a Fabry-Perot cavity containing the pink ruby for the input light nearly resonant to $R_1$ line (694 nm) with the input power level below 20 mW, resulting from a huge nonlinear refractive index [27]. A large amount of frequency chirping was observed in a Q-switched ruby laser operating at 694 nm [28], which would correspond to a *resonance-enhanced* $n_2$-value on the order of $10^{-16}$ m²/W under lasing conditions, in reference to values of various materials measured by the nonlinear transmittance measurement method [29].

Here, we examine the Kerr-lens effect based a nonlinear refractive index due to the dispersive contribution, which is expected to result from the population-density changes caused by high-intensity spikes, to provide new insights into self-induced spiking behaviors. The time-dependent focal length of a thin piece (thickness *d*) of a Kerr lens is given by

$$f_K = \pi w_e^4 / 8 n_2 d P_c(t), \tag{7}$$

where $w_e$ is the effective lasing beam spot size and $P_c(t)$ is the circulating intracavity laser power within the ruby crystal [30]. The focal length for the Kerr lens is determined by $P_c$ on the basis of the experimental peak power versus pump power relation shown in Fig. 2(b), assuming 1% transmittance of the output mirror, $M_2$, $w_e (\cong w_p) = 72$ mm, and $n_2 d = 10^{-18}$ m³/W.

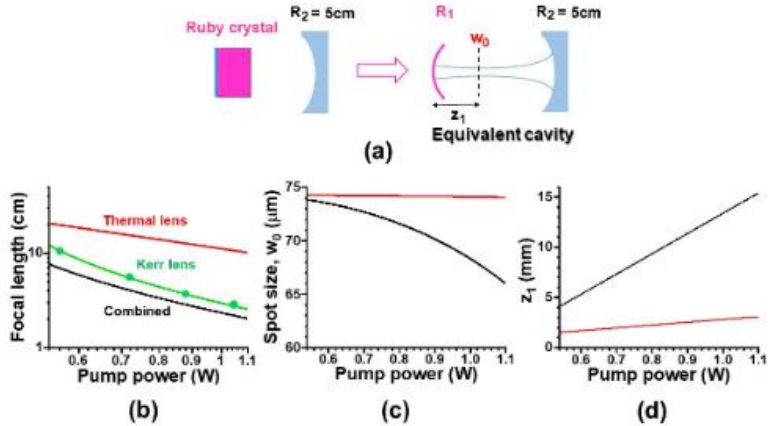

Fig. 5. (a) Equivalent laser-cavity configuration featuring thermal and Kerr lens effects. (b) Pump-dependent focal lengths of thermal, Kerr, and combined lenses. Pump and effective lasing beam spot sizes are assumed to be 72 ☐m. (c), (d) Pump-dependent beam waist spot size and its position within the cavity.

*4.2 Mode matching between pump and lasing beams*

In Fig. 5(b), the Kerr-lens effect shown by the green curve dominates the thermal-lens effect in the formation of the transverse eigenmode in the laser cavity. The equivalent laser cavity configuration incorporating thermal- and Kerr-lens effects is depicted in Fig. 5(a), where the effective radius of curvature of the coated end surface is given by $R_1 = 2f_C = 2[f_T \cdot f_K/(f_T + f_K)]$, with $f_C$ being the focal length of the combined lens. The resultant pump-dependent beam-waist spot size, $w_0$, and its position, $z_1$, given by the following equations are shown in Figs. 5(c) and 5(d), respectively.

$$w_0 = \sqrt{\frac{\lambda_o}{\pi}} \sqrt[4]{\frac{L(R_1 + R_2 - L)(R_1 - L)(R_2 - L)}{R_1 + R_2 - 2L}}, \tag{8}$$

$$z_1 = \frac{L(R_2 - L)}{R_1 + R_2 - 2L}. \tag{9}$$

The lasing mode profiles within the ruby laser crystal calculated using $w_0(z_1)$ values are shown in Fig. 6 for the thermal lens only (upper traces) and the combined lens incorporating the thermal and Kerr lens effects (lower traces), assuming $R_2 = 5$ cm and an optical cavity length, L = 2.5cm, together with the pump profile. The mode matching between the pump and laser beams is found to be greatly improved by the Kerr-lens focusing for the spiking mode, where the gain medium itself acts as a "soft aperture" for the cw mode in the terminology of Kerr-lens mode-locking.

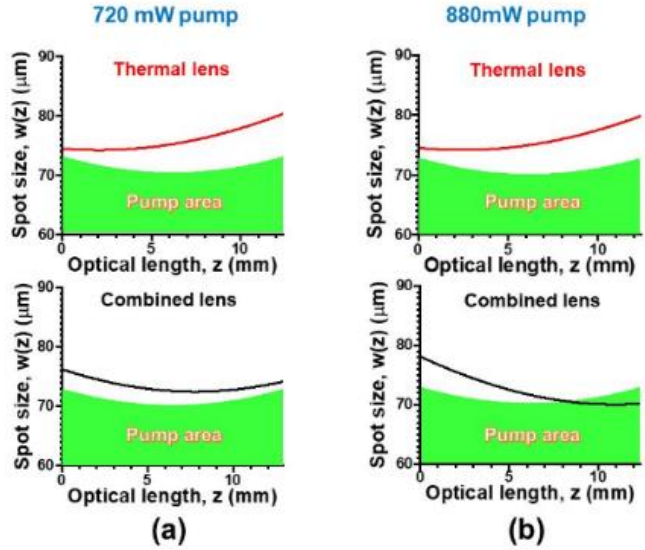

Fig. 6. Spot size variation along the ruby laser crystal with thermal lens (upper traces) and combined lens (lower trances)

On the other hand, the laser's slope efficiency, $S_e$, is shown to depend on *modal coupling efficiency*, $\eta_c$, which is determined by the three-dimensional overlap integral of the pump beam, r(x,y,z), and the laser beam, s(x,y,z), obeying the following equations [31]

$$\eta_c = \frac{(\iiint_{crystal} r(x,y,z)s(x,y,z)dv)^2}{\iiint_{crystal} r^2(x,y,z)s(x,y,z)dv}, \tag{10}$$

$$S_e = \frac{t_2 \nu_l \, \eta_c \eta_a}{L_c \nu_p} \tag{11}$$

Here, $L_c$ is a total cavity loss, $t_2$ is a transmittance of $M_2$, and $\eta_a$ is the pump absorption coefficient and beam intensities are normalized as

$$\iiint_{crystal} r(x,y,z)dv = 1, \quad \iiint_{crystal} s(x,y,z)dv = nl/L. \tag{12}$$

From these relations, the cavity loss decreases in proportion to the modal coupling efficiency given by Eq. (10). The equivalent time-dependent effective loss reduction associated with the time-dependent Kerr-lens effect (i.e., the lower traces in Fig. 6) is calculated to be $\delta L_c/L_c = 2.3 \times 10^{-2}$ (2.3%) and $5.1 \times 10^{-2}$ (5.1%) with respect to the thermal lens only (i.e., the upper traces in Fig. 6). Here, $\eta_c$ was calculated based on the experimentally determined beam profiles shown in Fig. 6(a) and Fig. 6(b). Such a dynamical loss reduction provides the laser with the possibility of encouraging spiking operation rather than cw operation through the Kerr-lens mediated balance of the gain and cavity loss, which includes the ground-state reabsorption in the unpumped region in Fig. 6.

A similar analysis to that of the periodic loss modulation in a laser Toda potential [17] shows that a spike-mediated driving force given by

$$F_D(t) = K\gamma\big(P'_c(t) + P_c(t)\big) \tag{13}$$

arises, as depicted by the arrow in Fig. 4(a), where $\gamma$ is determined in proportion to $n_2$-value in the form of $\gamma \propto 8n_2 d/w_e^4$. In the present ruby laser, the scaling factor is extremely large as $K = 1.7 \times 10^6$ and a pronounced driving force is expected to reduce dissipative resistive force as if the particle is moving through a viscous medium.

Note that the self-Q-switching operation based on the Kerr-lens effect was reported most recently in laser-diode-pumped Nd:LuAG lasers, where SQS was further verified to occur in a numerical simulation of the laser rate equations, including a similar photon-density-dependent intracavity loss to what we discussed above, considering soft aperture loss and self-focusing caused by the Kerr lens [32, 33].

## 5. Statistical analysis of self-induced instability in ruby laser

Finally, let us examine nonlinear dynamics, i.e., quasi-periodic spiking oscillations and chaotic state, observed outside the periodic spiking regime addressed in **2.2**. Time-dependent analyses of singular-value decomposition (SDV) spectra [34] for Fig. 3(b) are shown in Fig. 6(a), where the analysis was carried out in the time interval of a 4096- data length for each calculation with the shift of a 1024-data length for the next calculation. The steep exponential decay in the first segment followed by the noise floor clearly ensures the existence of chaos in the present ruby laser.

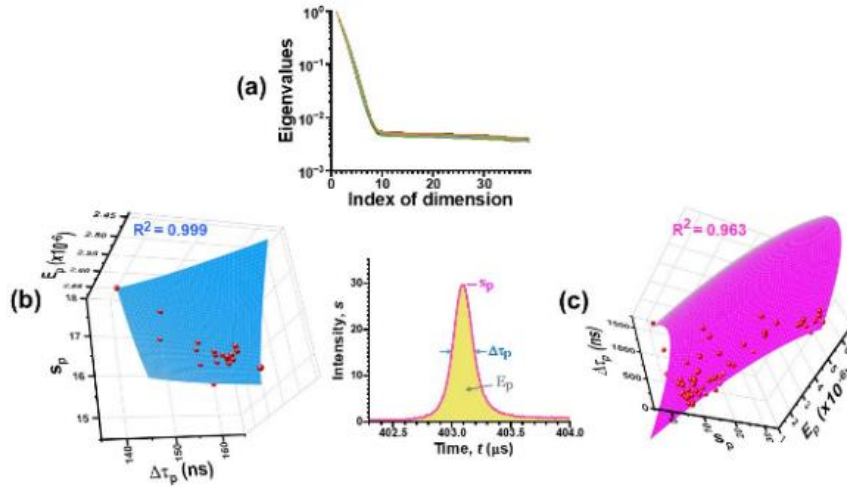

Fig. 7. (a) SDV spectra for chaotic state. 3D statistical graphics for (b) quasi-periodic spiking and (c) chaotic state. Fitting parameter values [$E_{p,0}$, a, b, c, d]: (b) [-5.0x10$^{-6}$, 2.91x10$^{-7}$, 35.26, -4.05x10$^{-9}$, -5.97x10$^{7}$]; (c) [-2.53x10$^{-6}$, 3.80x10$^{-7}$, 8.41, -4.48x10$^{-9}$, -4.02x10$^{6}$].

The intriguing relation between three quantities of peak intensity, pulse width, and pulse energy (area) of spikes is shown as the 3D statistical graphics in Figs. 7(b) and 7(c) for quasi-periodic spiking and chaotic state. Note that individual pulses with different energies are self-organized to lie on the 2-dimensional parabolic surface $E_p = E_{p,0} + a \cdot s_p + b \cdot \Delta\tau_p + c \cdot s_p^2 + d \cdot \Delta\tau_p^2$ for both cases. In particular, $R^2 = 0.999$ was obtained for quasi-periodic spiking oscillation.

In the case of chaotic self-spiking (hard mode) oscillations in solid-state lasers subjected to strong external modulation, a high degree of self-organization, with $R^2 = 0.996$, was clarified to be established behind the inverse-power-law universality of their intensity probability distributions [35, 17], while occasional interruptions of small-amplitude chaotic relaxation oscillations (soft mode) during chaotic spiking oscillations (hard mode) result in a breakup of the inverse power law, giving rise to the appearance of a peculiar "slope" [17]. On the other hand, the present ruby laser violates the inverse power law, as shown Fig. 8, where the peculiar "slope" appears to represent a quiet region of s(t) between the upper bound of the soft-mode chaos intensity fluctuations and the lower bound of the hard-mode chaotic fluctuations. Accordingly, self-organization is considered to be degraded to $R^2 = 0.963$.

Note that the intriguing statistical nature inherent to chaotic oscillations, which are brought about in solid-state lasers usually by external modulations, is present in the self-induced instabilities in the three-level ruby laser.

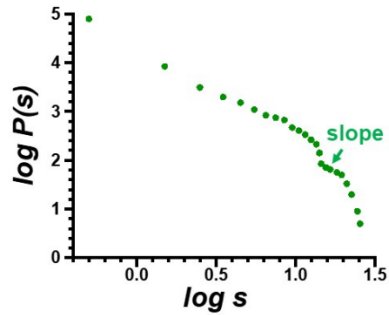

Fig. 8 Intensity probability distribution of chaotic state

## 6. Summary and outlook

In summary, self-induced spiking instability leading to chaos was demonstrated to occur in a ruby laser pumped with a continuous 532-nm laser. Dynamic behaviors in the form of periodic, quasiperiodic spiking oscillations and chaotic states were characterized by plotting waveforms, power spectra, and Poincaré sections. The intensity probability distribution of periodic spiking oscillations was found to result from the inherent asymmetric hyperbolic spiking waveform. Spiking oscillations were explained in terms of a particle moving in the laser Toda potential. Precise analyses of the experimental results revealed that the lasing cavity modes were formed by the pump-induced thermal lens effect for the cw mode and that high intensity spiking was associated with the dynamic Kerr-lens effect. The resulting enhanced mode-matching between the pump and lasing beams was described in terms of the mode coupling efficiency by computing the overlap integrals of the two interacting beams and the dynamics of the photon density were found to be analogous to a particle experiencing driving force while moving through a viscous medium.

The quasi-periodic spiking and chaotic dynamics observed at pump powers outside the periodic spiking regime in the ruby laser were characterized statistically. A time-dependent singular-value-decomposition (SVD) of an experimental time series showed that the chaotic spiking state appeared as the pump power increased. 3D statistical graphical analyses revealed an intriguing relation between the peak intensity, width, and energy of the spikes in the quasi-periodic and chaotic states, wherein individual pulse energies are self-organized so as to lie on a parabolic surface with a large coefficient of determination. Moreover, self-induced instabilities that are inherent to chaotic oscillations brought about in solid-state lasers usually by external modulation were found in the three-level ruby laser.

The results of the present study pose a crucial question regarding stable cw operation of ruby lasers: How do different dynamical behaviors (i.e., either non-spiking oscillation or self-induced spiking oscillation) take place within essentially the same laser cavity configuration but at different pump wavelengths, e.g., 514.5 nm, 405 nm, and 532 nm? To the best of our knowledge, this until-now hidden property of pink ruby crystals has not been reported in other solid-state lasers, including ones made from four-level or quasi-three-level materials. A possible answer could lie in the critical dependence of the dynamic stability on the relation between the pump and lasing beam profiles within the ruby crystal, reflecting the pump-wavelength-dependent absorption coefficient as well as the pump beam focus on the crystal.

**Acknowledgements:** The authors thank Prof. Jing-Yuan Ko, National Kaohsiung Normal University, Taiwan for his support in the SDV analysis.

**Disclosures:** The authors declare no conflicts of interest.

**Data availability:** Data underlying the results presented in this paper are not publicly available at this time but may be obtained from the authors upon reasonable request.